\documentclass[tradiabstract]{aa}

\newcommand{\bea}{\begin{eqnarray}}
\newcommand{\eea}{\end{eqnarray}}
\newcommand{\be}{\begin{equation}}
\newcommand{\ee}{\end{equation}}
\newcommand{\bc}{\begin{center}}
\newcommand{\ec}{\end{center}}
\newcommand{\ben}{\begin{enumerate}}
\newcommand{\een}{\end{enumerate}}
\newcommand{\bd}{\begin{description}}
\newcommand{\ed}{\end{description}}
\newcommand{\bmi}[1]{\begin{minipage}{#1 cm}}
\newcommand{\emi}{\end{minipage}}
\newcommand{\bmif}[1]{\begin{minipage}{#1\textwidth}}

\def\eck#1{\left\lbrack #1 \right\rbrack}

\def\rund#1{\left( #1 \right)}

\def\ave#1{\left\langle #1 \right\rangle}

\def\d{{\rm d}}

\def\arcsecf {\hbox{$.\!\!^{\prime\prime}$}}

\def\Real{{\rm I\mathchoice{\kern-0.70mm}{\kern-0.70mm}{\kern-0.65mm}%
  {\kern-0.50mm}R}}
\def\C{\rm C\kern-.42em\vrule width.03em height.58em depth-.02em
       \kern.4em}

\def\bx#1{\leavevmode\thinspace\hbox{\vrule\vtop{\vbox{\hrule\kern1pt
        \hbox{\vphantom{\tt/}\thinspace{\bf#1}\thinspace}}
      \kern1pt\hrule}\vrule}\thinspace}

\def\vc#1{{\mbox{\boldmath$#1$\unboldmath}}}
{\catcode`\@=11
\gdef\SchlangeUnter#1#2{\lower2pt\vbox{\baselineskip 0pt \lineskip0pt
  \ialign{$\m@th#1\hfil##\hfil$\crcr#2\crcr\sim\crcr}}}
}

\def\ueber#1#2{{\setbox0=\hbox{$#1$}%
  \setbox1=\hbox to\wd0{\hss$\scriptscriptstyle #2$\hss}%
  \offinterlineskip
  \vbox{\box1\kern0.4mm\box0}}{}}

\def\bx#1{\leavevmode\thinspace\hbox{\vrule\vtop{\vbox{\hrule\kern1pt
        \hbox{\vphantom{\tt/}\thinspace{\bf#1}\thinspace}}
      \kern1pt\hrule}\vrule}\thinspace}

\def\arcsecf {\hbox{$.\!\!^{\prime\prime}$}}

{\catcode`\@=11
\gdef\SchlangeUnter#1#2{\lower2pt\vbox{\baselineskip 0pt \lineskip0pt
  \ialign{$\m@th#1\hfil##\hfil$\crcr#2\crcr\sim\crcr}}}
}

\def\ts{\thinspace}

\usepackage{graphicx}
\usepackage{txfonts}
\usepackage{natbib}
\usepackage{subfigure}
\usepackage{enumerate}
\usepackage{slashbox}
\begin{document}

   \title{Mass-sheet degeneracy, power-law models and external
     convergence: Impact on the determination of the Hubble constant
     from gravitational lensing}

   \author{Peter Schneider \inst{1} \& Dominique Sluse
          \inst{1}
          }

   \institute{Argelander-Institut f\"ur Astronomie, Universit\"at
     Bonn, Auf dem H\"ugel 71, D-53121 Bonn, Germany\\
    peter, dsluse@astro.uni-bonn.de}



  \abstract
  { The light travel time differences in strong gravitational lensing
    systems allows an independent determination of the Hubble
    constant. This method has been successfully applied to several
    lens systems. The formally most precise measurements are, however,
    in tension with the recent determination of $H_0$ from the Planck
    satellite for a spatially flat six-parameters $\Lambda CDM$
    cosmology. We reconsider the uncertainties of the method,
    concerning the mass profile of the lens galaxies, and show that
    the formal precision relies on the assumption that the mass
    profile is a {perfect} power law. Simple analytical
      arguments and numerical experiments reveal that mass-sheet like
    transformations yield significant freedom in choosing the mass
    profile, even when exquisite Einstein rings are
    observed. Furthermore, the characterization of the environment of
    the lens does not break that degeneracy which is not physically
    linked to extrinsic convergence. We present an illustrative
    example where the multiple imaging properties of a composite
    (baryons + dark matter) lens can be extremely well reproduced by a
    power-law model {having the same velocity dispersion}, but
    with predictions for the Hubble constant that deviate by $\sim
    20\%$.  Hence we conclude that the impact of degeneracies between
    parametrized models have been underestimated in current $H_0$
    measurements from lensing, and need to be carefully
    reconsidered. }

  \titlerunning{$H_0$ from gravitational lensing}

   \maketitle
%

\section{Introduction}
Half a century ago, Sjur Refsdal (\citeyear{Refsdal1964}) pointed out
that a gravitational lens system can be used for determining the
Hubble constant $H_0$, provided that the difference in light-travel
time along the different light rays can be measured. The
identification of active galactic nuclei as variable, distant
extragalactic sources, as well as the discovery of gravitational lens
systems allowed this idea to be turned into a target of
research. Immediately after the discovery of the first gravitational
lens system (Walsh et al.\ \citeyear{wcw79}), the flux variations of
its two compact components were monitored, and a firm measurement of
the time delay was obtained in 1997 (Kundic et al.\
\citeyear{kundic97}), with sub-percent accuracy. Since then, the time
delays in about 20 other lens systems have been measured
\citep[see e.g. ][for a compilation and recent
measurements]{Paraficz2010, Courbin2011, Eulaers2011, Tewes2012,
Fohlmeister2013, Eulaers2013}.

To transform a measured time delay into an estimate of the Hubble
constant, the mass distribution in the lens system needs to be
determined with sufficient accuracy. Since our physical understanding
of the mass distribution in the inner parts of galaxies is
insufficient, parametrized mass models are employed whose parameters
are fixed by the observational constraints, like the angular positions
of compact images of the source, morphology of extended image
components, or sometimes flux ratios. Reassuringly, `simple'
parametrized mass models yield an accurate description of the image
morphologies in most lens systems, without the need for more complex
matter distributions. These models are further constrained by
additional information, such as measurements of the stellar velocity
dispersion in the lens galaxy.  As a result, one obtains a good model
(in the sense that it fits the data) for the mass distribution in the
inner part (i.e. within a few effective radii) of lens galaxies, which
together with time delay 
measurements is used to estimate $H_0$.

An excellent account of the early history of $H_0$ measurements from
lensing can be found in \cite{koch2004}. In particular, estimated
values of $H_0$ varied substantially from system to system, and even
different analyses for the same system sometimes gave substantially
differing values. This can be traced back to the non-uniqueness of the
mass model.  Since a typical gravitational lens system has two or four
images, the number of observational constraints on the mass model is
small, and there is more than one density profile which can reproduce
the observed image positions. Different values of $H_0$ result if one
assumes either an isothermal profile (one where the volume mass
density behaves like $\rho \propto r^{-2}$) or a mass profile which
follows closely the projected brightness profile. These two different
model assumptions lead to different slopes of the density in the
region where the strong lensing constraints are available, and this
slope affects the product of time delay and Hubble constant, $\tau =
H_0\,\Delta t$ \citep{KOC2002}.  Indeed, \cite{FGS85} pointed out the
so-called mass-sheet degeneracy (MSD), which provides a transformation
of the mass profile which leaves all lensing observables exactly
invariant, except $H_0\,\Delta t$. The mass-sheet degeneracy and
families of degeneracies have been discussed in \cite{Gorenstein1988,
  Saha2000, Wucknitz2002, Liesenborgs2012}. The MSD is the basic
reason why the estimated 
values of $H_0$ from lensing {{have been}} so different.

More recently, Suyu et al. (\citeyear{Suyu2010, Suyu2013a}) studied
two four-image lens systems in great detail, making use of multi-color
images from the Hubble Space Telescope and spectroscopy of the lens
galaxies in these systems. Together with the measured time delays,
they obtained a value of $H_0=70.6\pm3.1\, {\rm km\, s^{-1} Mpc^{-1}}$
for B1608+656 and $H_0=78.7^{+4.3}_{-4.5}\, {\rm km\, s^{-1}
  Mpc^{-1}}$ for RXJ1131--1231.  Motivated by the recent results from
the Planck satellite (Planck Collaboration \citeyear{PLXVI}), which
yield a value of $H_0=67.3\pm 1.2 \,{\rm km\, s^{-1} Mpc^{-1}}$ for a
spatially flat $\Lambda$CDM cosmology, we reconsider some issues
related to the determination of the Hubble constant from gravitational
lensing. In Sect.\ts 2, we briefly review the basics of the MSD and
its consequences. We then discuss in Sect.\ts\ 3 the uniqueness of
power-law mass models, which have been widely employed for strong lens
modeling, and the impact of possible deviations from a power law. More
complicated (though arguably more realistic) mass profiles are
considered in Sect.\ts 4. We show that those models are (almost)
degenerate with global power-law profiles leading to quite different
values of $\tau$.  Sect.\ts 5 considers the original interpretation of
the mass-sheet degeneracy, as an additional surface mass density
related to a galaxy cluster, or some overdensity along the
line-of-sight. We will argue that the mass-sheet degeneracy is in
general not broken by observing the environmental (or `external')
convergence. We briefly conclude in Sect.\ts\ 6.

\section{The mass-sheet degeneracy}
We use standard gravitational lensing notation \citep[][]{SEF}; in
particular, 
$\kappa(\vc\theta)$ is the dimensionless surface mass density (or
convergence) at angular position $\vc\theta$, $\psi(\vc\theta)$ the
deflection potential, satisfying the two-dimensional Poisson equation
$\nabla^2\psi=2\kappa$, and $\vc\alpha(\vc\theta)=\nabla\psi$ is the scaled
deflection angle. The lens equation
$\vc\beta=\vc\theta-\vc\alpha(\vc\theta)$ relates the observed angular
position $\vc\theta$ on the sky to the true intrinsic position
$\vc\beta$ in the absence of light deflection. The Jacobian matrix of
the lens mapping is denoted by $\cal A(\vc\theta)$, and the (signed)
magnification of an image of an infinitesimally small source is 
$\mu=1/\det{\cal A}$. Critical curves in the lens plane, i.e., curves
of formally infinite magnification, are characterized by $\det{\cal
  A}=0$. 

\subsection{The mass-sheet transformation}

As shown by Falco et al.\ (\citeyear{FGS85}), the mass distribution
$\kappa(\vc\theta)$ and each of the distributions
\be
\kappa_\lambda(\vc\theta)=\lambda\,\kappa(\vc\theta)+(1-\lambda)\;,
\label{eq:MSD}
\ee
together with an (in most cases unobservable; see below) isotropic
scaling of the source plane coordinates $\beta \to \lambda\,\beta$,
yields exactly the same dimensionless observables, i.e., image
positions, image shapes, magnification ratios, etc. This is called the
mass-sheet degeneracy (MSD).  In other words, from the observed image
positions and flux ratios, one cannot distinguish between the original
$\kappa$ and any of the mass distributions in (\ref{eq:MSD}). As shown
by \cite{ScCS95}, also weak gravitational lensing
cannot break the MSD, since image shapes are unaffected.  However, the
product of the time delay and the Hubble constant is affected,
$H_0\,\Delta t \to \lambda\, H_0\,\Delta t$, but leaving time
delay ratios again invariant.

\subsection{Breaking the mass-sheet degeneracy}
The mass-sheet transformation (MST) leaves the critical curves
invariant, also the curves on which $\kappa=1$. Furthermore, it leaves
the shapes of the isodensity contours invariant, just the value of
$\kappa$ on these contours changes according to (\ref{eq:MSD}). 

As is clear from the transformation of $H_0\,\Delta t$, in order to
get a reliable estimate of the Hubble constant from gravitational
lensing, one first needs to break the MSD.  Several ways have been
suggested in the literature. Some of these make use of the fact
that the MST (\ref{eq:MSD}) affects the magnification, $\mu \to
\mu/\lambda^2$, hence if the magnification can be estimated, the value
of $\lambda$ can be constrained (this magnification corresponds to the
aforementioned isotropic scaling of the source plane coordinates).
For AGN as sources, which have a very broad distribution of intrinsic
luminosities, this cannot be easily accomplished. Whereas
\cite{Bauer12} have shown that the correlation between AGN variability
properties and luminosity can be used as a tool for estimating source
luminosities, and hence lensing magnification, the scatter of the
variability--luminosity relation is large and can only be employed in
a statistical way.

Another possibility to break the MSD in strong lensing systems is
based on independent mass estimates of the lens. Combining lensing
measurements with spectroscopy of the lens galaxy, the MSD can be
broken. {The use of velocity dispersion measurements in estimates
  of $H_0$ with the time-delay technique started with the modeling of
  Q0957+561 by \cite{Grogin1996}. It was later refined by
  \cite{Romanowsky1999} who suggested to use higher-order moments of
  the stellar velocity distribution.}  For early-type galaxies (most
lenses are of that type), the stellar velocity dispersion yields an
estimate of the mass inside the effective radius of the lens, which
together with the precise (and unaffected by the MST) determination of
the mass inside the Einstein radius of the lens allows one to
determine the mean slope $\gamma'$ of the mass profile between the
effective radius and the Einstein radius. {{The Lens Structure and
    Dynamics Survey (LSD) initiated the use of velocity dispersion to
    derive the average slope of the mass density profile of lensing
    galaxies~\citep{Treu2002}.}}  Since then, many more systems have
been studied \citep{Rusin2005, Koopmans2009}. In particular,
\cite{Koopmans2009} derived an average slope $
\gamma'=2.085^{+0.025}_{-0.018}$ (with an intrinsic scatter
$\sigma_{\gamma'} \sim 0.2$) for 53 early-type galaxies discovered by
the Sloan Lens ACS survey (SLACS). {Refinements of this technique
  are used in modern measurements of $H_0$ with lensing (Suyu et
  al. \citeyear{Suyu2010, Suyu2013a}).}

The velocity dispersion and
lensing constraints have generally been treated
independently. \cite{Barnabe2007} developed a unified scheme to
retrieve a gravitational potential which reproduces the observed
stellar dynamics of the lens (instead of the velocity dispersion only)
and gravitational lensing observables simultaneously. This technique
has been applied to several lensing galaxies from the SLACS sample
\citep[e.g.][]{Barnabe2009, Barnabe2011}. {{The measurement of
    $H_0$ from time-delay lenses currently makes use only of the
    central velocity dispersion information.}}

In general, the velocity dispersion measurement in individual systems
is derived with typically 10\% uncertainty, which translates into an
uncertainty of the same order on the logarithmic slope of the profile
\citep[e.g.][]{Koopmans2004, Auger2010b, Agnello2013}. {{Because
    the relative uncertainty on $H_0$ is at least equal to the
    relative uncertainty on the slope \citep[e.g.][]{Refsdal1994,
      Witt1995, Witt2000, KOC2002}, the velocity dispersion
    measurement is of little help in reducing the uncertainty on $H_0$
    below 10\%. In their analysis of B1608+656 and J1131$-$1231, Suyu
    et al.\ (\citeyear{Suyu2010}, \citeyear{Suyu2013a}) have derived
    an uncertainty on the slope of the mass distribution of $<3\%$ while
    their uncertainty on the velocity dispersion is about 6\%. Hence,
    the latter measurement contributes only little in reducing the
    formal uncertainty on the slope. Furthermore,}} the
radial/tangential anisotropy of the stars commonly encoded in the
anisotropy parameter $\beta = 1-(\bar{\sigma^2_{\theta}}/\bar
{\sigma^2_{r}})$ also systematically affects the estimate of the slope
to a level which can reach 15\% \citep[see Fig. 1 of][]{Koopmans2009}.
{\cite{Agnello2013}} argue that the impact of anisotropy may in
practice be smaller, i.e. less than 5\%.

{In any case, the MST to first order corresponds to a scaling of
  the three-dimensional mass distribution by a factor $\lambda$ --
  with the constant $(1-\lambda)$-term corresponding to a larger-scale
  3-D mass component which contributes little to the gravitational
  potential inside the effective radius. Since $\sigma^2\propto M$, we
  find that $\Delta H_0/H_0=\Delta\lambda/\lambda
  =\Delta M/M=2\Delta\sigma/\sigma$. Thus an uncertainty of $6\%$ in
  the stellar velocity dispersion translates into a $\sim 12\%$
  uncertainty in the Hubble constant, even if we ignore uncertainties
  regarding orbit anisotropies and triaxiality of the mass
  distribution. Therefore, the accuracy of $H_0$ claimed in Suyu et
  al.\ (\citeyear{Suyu2010}, \citeyear{Suyu2013a}) is not due to MSD
  breaking by stellar dynamical information.}

\section{Power-law mass profiles}
The surveys just mentioned have demonstrated that the mean slope of
the three-dimensional density profile of early-type lens galaxies,
measured between the effective radius and the Einstein radius, is
nearly isothermal, i.e.  $\gamma'\sim 2$, with relatively small
variations ($\sim\pm 0.2$) between different lenses. It must be
pointed out that the slope of the density profile determined by this
method is an average one, over the radial range between the effective
radius and the Einstein radius; it does not imply that the density
profile is indeed accurately described by a power law. In fact, from
our understanding of the physical properties of galaxies, one would
not expect the density profile {{in the central kiloparsecs region
    where the lensing effects occur}} to be a {{perfect}} power
law \citep[e.g.][but see also Remus et
al. \citeyear{Remus2013}]{vandeven2009, Capellari2013}. These
aforementioned surveys have concluded that about half of the mass
inside the Einstein radius is contributed by baryonic material, the
other half by dark matter (somewhat dependent on the assumed initial
mass function, and thus the mass-to-light ratio, of the stellar
population). Neither the luminous matter, nor the dark matter,
individually are expected to follow an isothermal profile: The light
(and baryonic mass) distribution in early-type galaxies is well
described by a S\'ersic profile, whereas dark matter-only numerical
simulations suggest the dark matter in halos to have a `universal'
profile, with an inner slope close to $\gamma'\sim 1$
\citep{NFW96}. The dark matter density profile of real galaxies
probably deviates from this universal profile, due to feedback
processes and the contraction of the dark matter halo through the
condensation of baryons due to cooling in the inner part. In any case,
the fact that the observed mean slope is close to isothermal is most
likely a conspiracy and does not imply that the power-law assumption
can be extrapolated to radii smaller (or larger) than the Einstein
radius \citep{vandeven2009}.

Nevertheless, it has been common practice to model the radial mass
profiles of lenses by a power law (e.g. Rusin et
al. \citeyear{Rusin2003}, Suyu et al. \citeyear{Suyu2010, Suyu2013a}).
It must be pointed out that this assumption formally breaks the MSD,
since the transformation of a power law of $\kappa$ according to
(\ref{eq:MSD}) is {\it not} a power law. In other words, the
assumption of a power law picks out one member of the family
(\ref{eq:MSD}) of mass models. Acknowledging the fact that we have no
a priori reason to suspect $\kappa$ to be a true power law, this
formal breaking of the MSD may lead to biased results.

On the other hand, whereas a mass-sheet transformed power law is no
longer a power law, it is approximately so, as long as $\lambda$ is not
very different from unity. For a power-law mass distribution, one
finds that 
\be
\kappa\rund{\sqrt{\theta_1 \theta_2}}
=\sqrt{\kappa(\theta_1)\kappa(\theta_2)}\;.
\ee
If we now denote with $\theta_{\rm min}$ and $\theta_{\rm max}$ the
inner and outer radial coordinate of the region where strong lensing
constraints are available, the ratio
\be
\xi={ \kappa\rund{\sqrt{\theta_{\rm min} \theta_{\rm max}}}
\over \sqrt{\kappa(\theta_{\rm min})\kappa(\theta_{\rm max})} }
\label{eq:xi}
\ee
is a measure of the deviation from a power law, with $\xi=1$ for an
exact power law.  

\begin{figure}
\includegraphics[width=\hsize]{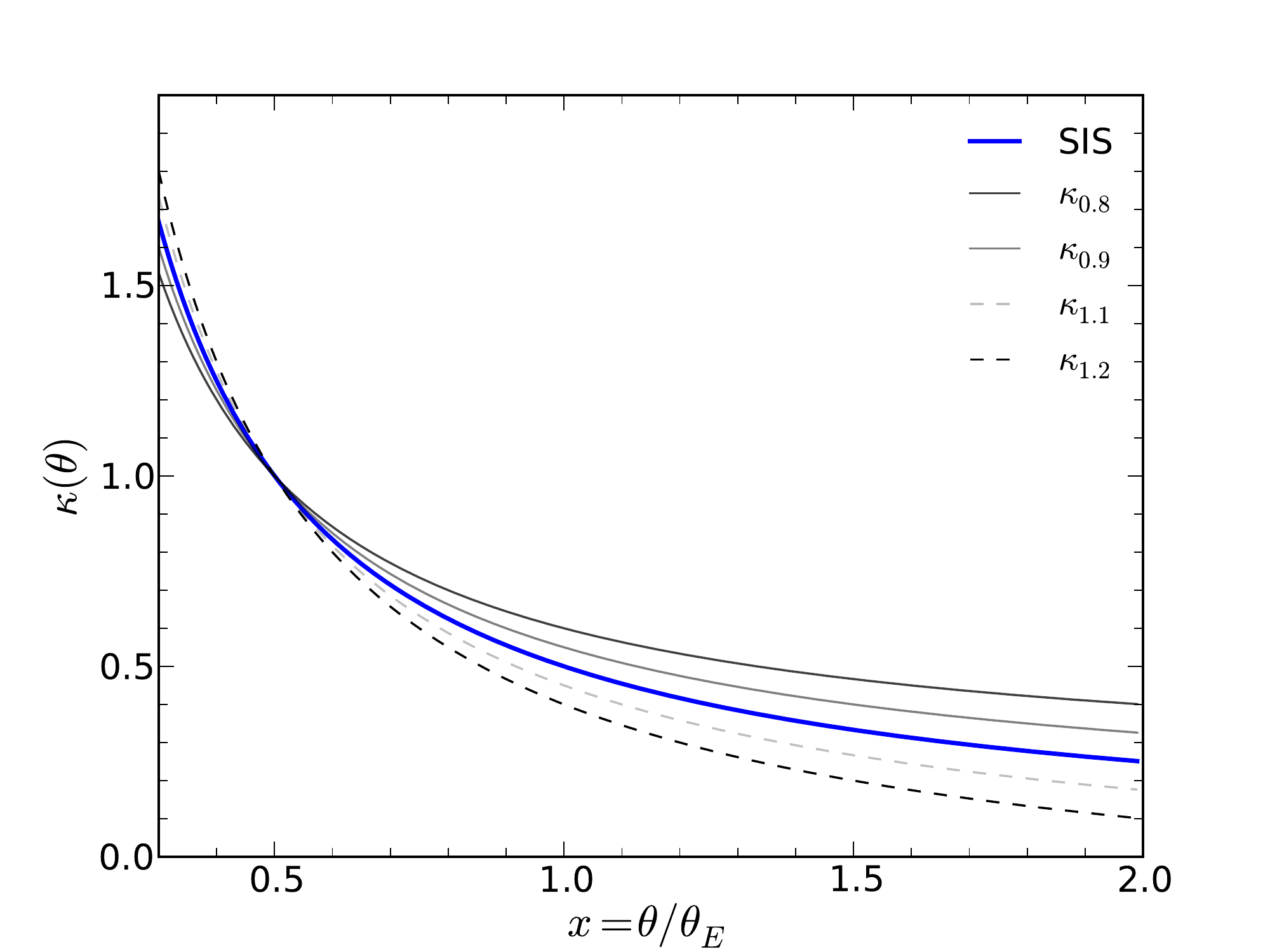}
\caption{The isothermal density profile (thick curve), and several of
  the transformed profiles $\kappa_\lambda(\theta)$, with $\lambda$
  ranging from 0.8 (flattest curve) to 1.2 (steepest curve), plotted
  as a function of $x=\theta/\theta_{\rm E}$. Note
  that, if the SIS density profile provides a good fit to the lensing
  data, an equally good fit is obtained by all the $\kappa_\lambda$}
\label{fig:1}
\end{figure}

We consider an isothermal model as reference, with $\kappa(\theta)=
\theta_{\rm E}/(2 \theta)$, where $\theta_{\rm E}$ is the Einstein
angle of the lens. In Fig.\ts\ref{fig:1} we plot the density profile
of a singular isothermal sphere (SIS), and several of the
$\kappa_\lambda(\theta)$ obtained from (\ref{eq:MSD}). We see that
over the radius range plotted, the transformed profiles appear almost as
power laws, with a slope depending on $\lambda$. We quantify the slope
in two different ways: First, by taking the local slope 
\be
s=-{\d \ln \kappa_\lambda\over \d \ln\theta}
\ee
at the Einstein radius, yielding $s=\lambda/(2-\lambda)$. Second, we
take the mean slope over the angular range $\theta_{\rm min}\le\theta
\le \theta_{\rm max}$,
\be
\bar s={ \ln\eck{ \kappa_\lambda(\theta_{\rm max})/
\kappa_\lambda(\theta_{\rm min})} \over
\ln(\theta_{\rm min}/\theta_{\rm max})}\;.
\label{eq:bars}
\ee
For an SIS lens, the image positions are separated by $2\theta_{\rm
  E}$; hence, if the positive parity image is found at
$\theta=x\,\theta_{\rm E}$, with $1<x<2$, the negative parity image is
located on the other side of the lens center, with
$|\theta|=(2-x)\theta_{\rm E}$. Thus we choose $\theta_{\rm
  min}=(2-x)\theta_{\rm E}$, $\theta_{\rm max}=x \theta_{\rm E}$.

\begin{figure}
\includegraphics[width=\hsize]{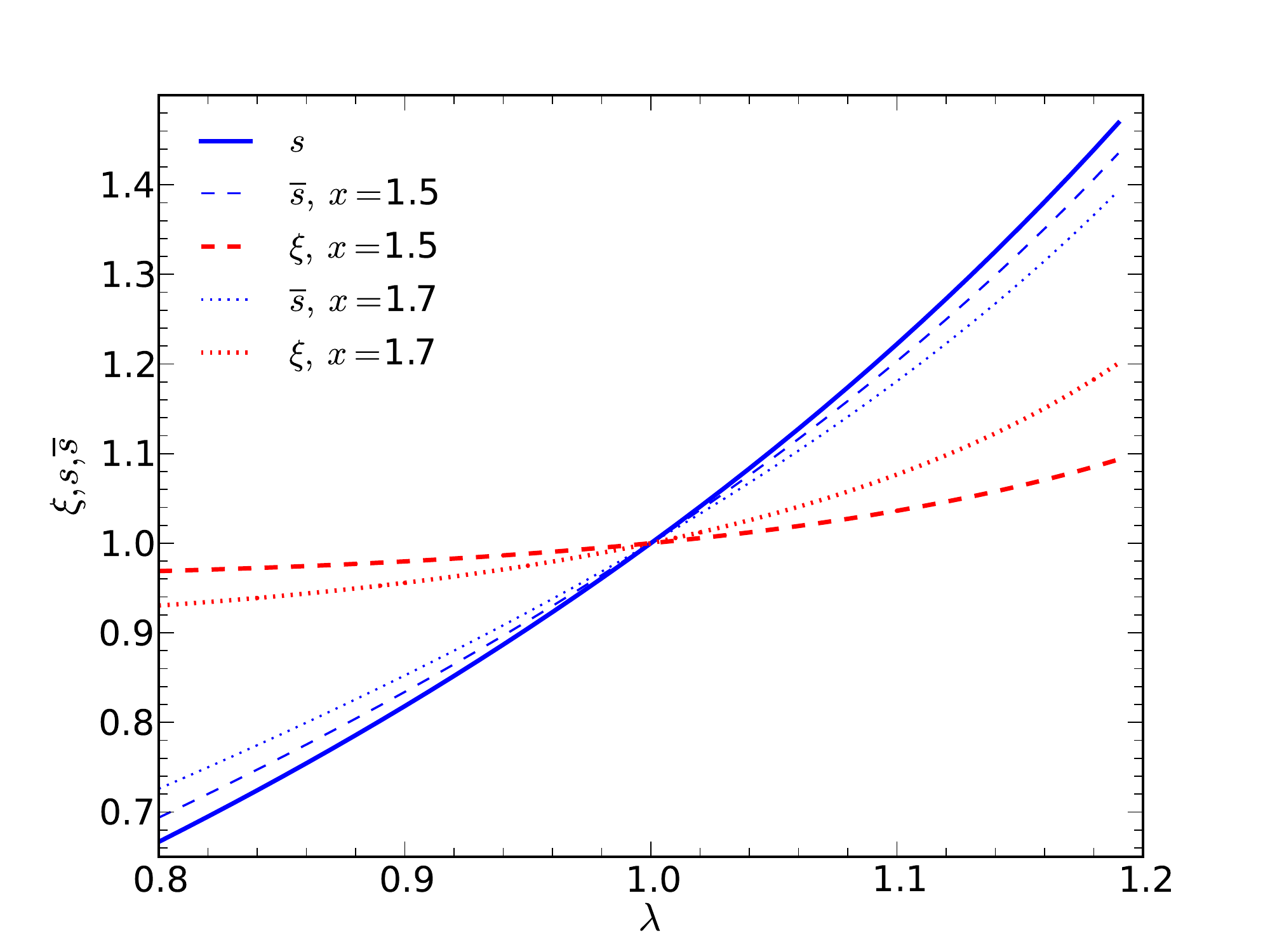}
\caption{For the mass models obtained by the transformation
  (\ref{eq:MSD}) of an SIS model, we plot the local slope
  $s=\lambda/(2-\lambda)$ at the
  Einstein radius as thick blue solid curve. The dashed and dotted
  thin blue curves show the
  mean slope $\bar s$  over the interval $(2-x)\theta_{\rm E} \le
  \theta \le x \theta_{\rm E}$, for $x=1.5$ and $x=1.7$,
  respectively. For the same 
  parameters, the dotted and dashed red thick curves show the quantity
  $\xi$ which provides 
  a measure of the deviation of $\kappa_\lambda$ from a power law,
  over the same angular range}
\label{fig:2}
\end{figure}

In Fig.\ts\ref{fig:2}, we have plotted the local slope $s$ at the
Einstein radius (solid blue curve), and the mean slope over the
intervals $0.3\le\theta/\theta_{\rm E}\le 1.7$ (dotted) and
$0.5\le\theta/\theta_{\rm E}\le 1.5$ (dashed) as thin blue curves, as
a function of $\lambda$. Furthermore, we plot the parameter $\xi$ (see
Eq.\ts\ref{eq:xi}) over the same intervals as thick red curves. We see
that $\xi$ stays close to unity over a fairly wide range of $\lambda$,
i.e., deviations from a power law are small over the intervals
probed. This can also be seen by the fact that the mean slope $\bar s$
is very similar to the local slope $s$. Hence, a mass-sheet
transformed power law remains a power law, to a good approximation
{{($\xi$ deviates from 1 by less than 5\% for $\lambda \in
    [0.8,1.1]$ for the case $x=1.5$)}}, over the radius range
typically probed by strong lensing, for quite a range of
$\lambda$-values.

\section{A composite model}

In the previous section, we have studied the MST of a power law. Here,
we present analytical and numerical experiments showing that the
astrometry and photometry of synthetic lensed systems resulting from
the lensing by the sum of two different density profiles can be
reproduced with a single power law profile. We show that this
degeneracy between profiles is coincidentally very similar to the
effect of a MST and impacts only the time delay.

We start in Sect.~\ref{subsec:toy} with a toy model which 
{provides a simplistic analytic illustration of}
the freedom one has in choosing the mass models for a circular lens
potential. In Sect.~\ref{subsec:2D}, we investigate numerically the
situation for a {more realistic}
non-circularly symmetric lens, and a complex and
structured source composed of 10 compact structures leading to more
than 20 lensed images (doubles or quadruples).

\subsection{Toy data set}
\label{subsec:toy}

We follow \cite{Suyu2012} who considered the possibility to constrain
the slope of lenses in the case of a two-image system. 

Hence, we assume to have a source consisting of two closely spaced
components. {For simplicity, we consider axi-symmetric mass
  distributions.  We furthermore assume that the lensing properties
  are such that the images of the source components are located at
  $\theta_1$ and $\theta_1+\Delta\theta$ on one side of the lensing
  galaxy, and at $\theta_2$ and $\theta_2-\Delta\theta$ on the other
  side of the lens center, with $\theta_1>\theta_2>0$.} Hence, the
separation of the two subcomponents is the same in both images. This
implies that, when we try to model this image system with a power-law
mass distribution, its slope will be isothermal. The Einstein radius
of the SIS is $\theta_{\rm E}=(\theta_1+\theta_2)/2$.

The same data can also be fit with a composite model, consisting of a
S\'ersic profile plus a power law, which should resemble a toy model of
the mass distribution of baryons and the local distribution of dark
matter in the inner part of the galaxy, respectively. Further
specializing the S\'ersic profile to an exponential, we write
\be
\kappa(\theta)=a \exp\rund{-{\theta\over\theta_{\rm d}}}
+b \rund{\theta\over\theta_{\rm E}}^{1-\gamma_{\rm DM}}\;,
\label{eq:compo}
\ee
where $\theta_{\rm d}$ is the scale radius of the exponential. For a
given choice of $\theta_{\rm d}$ and 3D-slope $\gamma_{\rm DM}$, one can find the
values of the amplitudes $a$ and $b$ which satisfy the two constraints
from the lensing information, i.e., which yield the same source
positions for the two multiply imaged source components. Hence, this
mass model can yield an exact solution of the lens equation. The
corresponding algebra does not yield useful insight and will not be
reproduced here.

\begin{figure}
\includegraphics[width=\hsize]{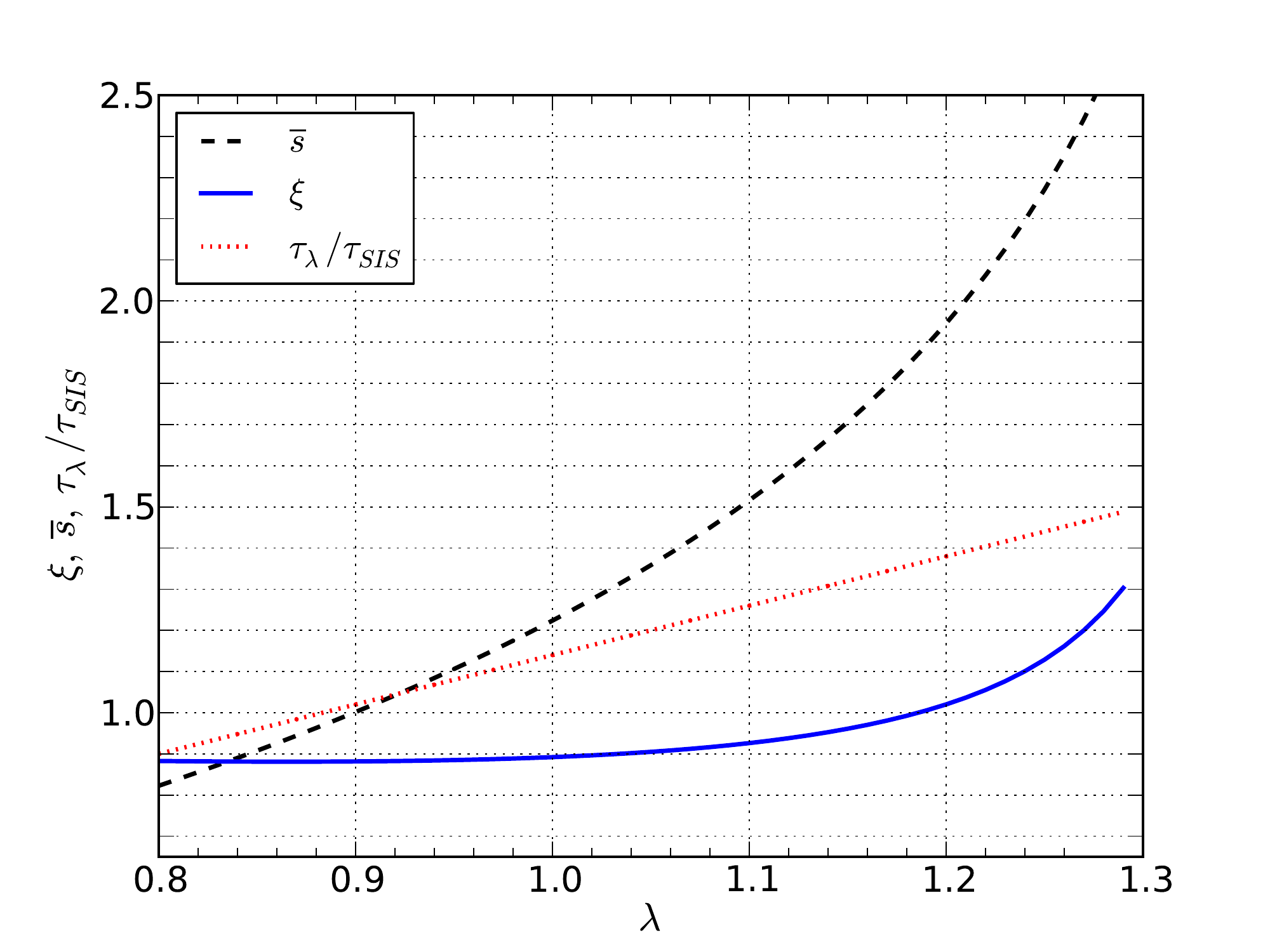}
\includegraphics[width=\hsize]{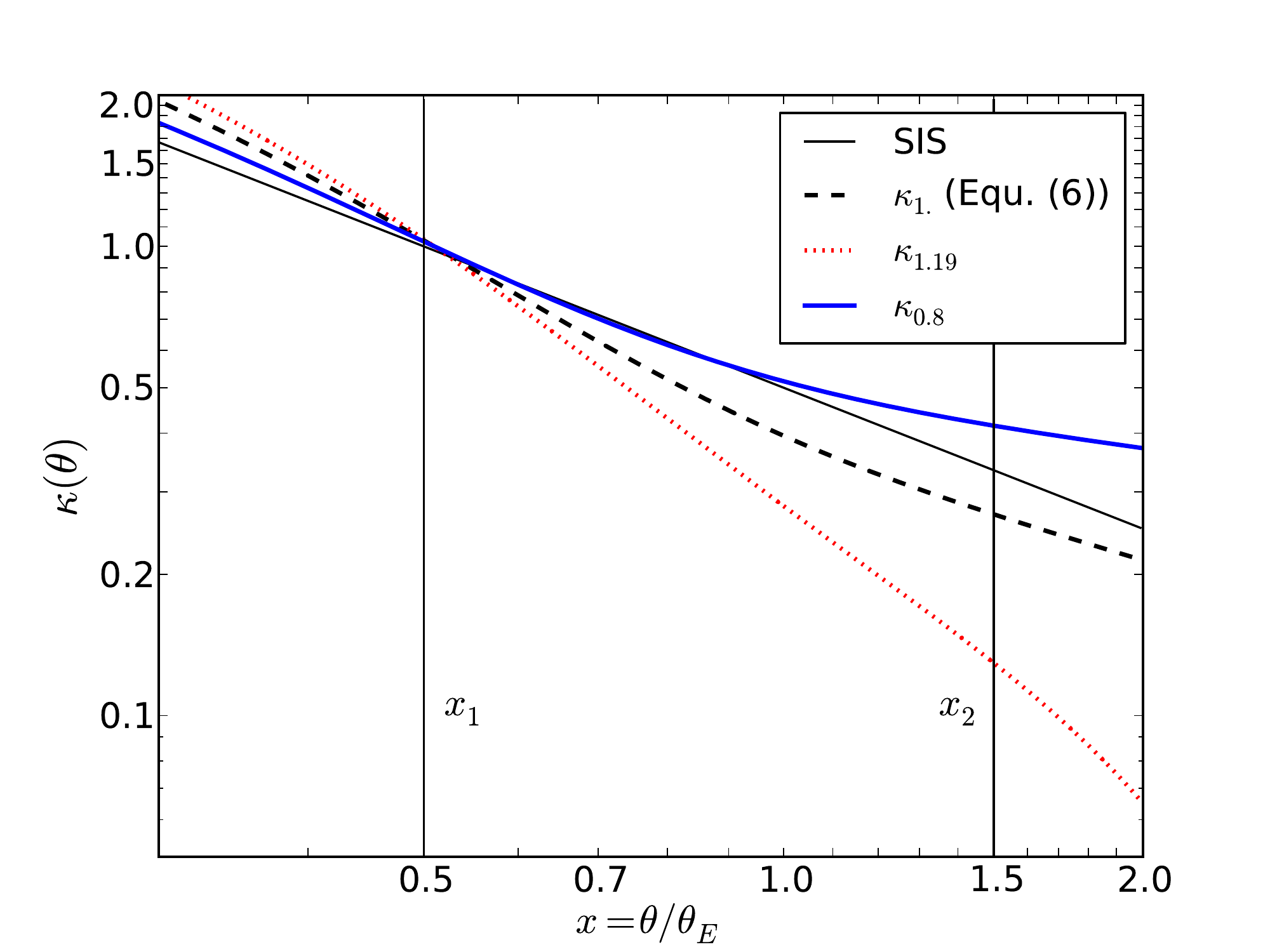}
\caption{Upper panel: The quantity $\xi$ (blue solid curve), the mean
  slope $\bar s$ (dashed black curve) and the ratio of the time delays
  $\tau = H_0\,\Delta t$ (dotted red curve) for the mass models
  $\kappa_\lambda$ and that of the singular isothermal sphere, as a
  function of $\lambda$. Lower panel: The mass profiles of the model
  described by (\ref{eq:compo}) (dashed black curve), two mass-sheet transformed
  ones ($\kappa_{0.8}$ -- solid blue curve, and $\kappa_{1.19}$ --
  dotted red curve), and that of the corresponding SIS (thin black
  curve)}
\label{fig:3}
\end{figure}

As an example, we choose $\theta_{1}=3\theta_2$, and
$\Delta\theta=\theta_{\rm E}/50$. Furthermore, we set
$\theta_{\rm d}=0.2 \theta_{\rm E}$ and
$\gamma_{\rm DM}=1.75$, yielding $a\approx 5.17$ and $b\approx0.36$. The
corresponding mass profile is shown as dashed black curve in the
bottom panel of Fig.\ts\ref{fig:3}; compared to the SIS model (shown
in the same panel as thin solid curve), it is slightly steeper within
the strong lensing region.
The corresponding time delay is about 15\% higher than for the SIS
model, as expected from the steeper slope.

Next, we consider mass-sheet transformed versions of the profile
described in (\ref{eq:compo}).  In the top panel of
Fig.\ts\ref{fig:3}, we show the parameter $\xi$ defined in
(\ref{eq:xi}), evaluated between the radii $x_1\equiv
\theta/\theta_{\rm E}=0.5$ and $x_2=1.5$, i.e., over the region where
strong lensing constraints are available, as a function of $\lambda$
(blue solid curve). Over most of the parameter range plotted in the
figure, $|\xi-1|\lesssim 0.1$, i.e., in the radial range probed by
strong lensing, the mass distribution fairly closely resembles a power
law.  The mean slope $\bar s$ of the mass profile between these two
radii is shown as dashed black curve; increasing $\lambda$ yields
steeper slopes, as expected. The red dotted curve shows the ratio of
the time delay of the composite model, compared to that of the SIS;
this ratio is a linear function of $\lambda$, as expected from the
MST.

In the bottom panel of Fig.\ts\ref{fig:3}, the resulting mass profiles
$\kappa_\lambda$ are shown for $\lambda=0.8$ (blue solid curve), the
original one $\kappa_1\equiv\kappa$ (black dashed curve) and for
$\lambda=1.19$ (red dotted curve), the value of $\lambda$ for which
$\xi\approx 1$, i.e., where the resulting mass profile is closest to a
power law in the strong lensing region. Hence we see that the same
strong lensing data can be fit with two different power laws, one
global one with isothermal slope, and one which is locally almost
exactly a power law, but with very different slope (and a resulting
37\% larger value of $\tau = H_0\,\Delta t$). 

We conclude from this very simple example that there is a large
freedom of fitting strong lensing data, not only due to the MSD, but
also to our lack of well-motivated shapes of the total mass
distribution of galaxies. Of course, this simple example may not be
very realistic; in particular, we have not studied whether this model
is able to satisfy any stellar kinematic constraints obtained from
spectroscopy of the lens galaxy. We expect that such a more detailed
study would yield constraints on the parameters $\theta_{\rm d}$ and
$\gamma_{\rm DM}$ in our model. However, the main conclusion -- a substantial
freedom in the choice of possible mass profiles for fitting lensing
data -- will be unchanged.

The profiles shown in the bottom panel of Fig.\ts\ref{fig:3} show a
substantially different behavior for larger radii, compared to the
SIS, and may appear unphysical. This is due to the simplified model of
the dark matter profile as a global power law. However, as we will
discuss is Sect.~\ref{sec:kappaext}, this behavior at large radii is
of no particular concern.

\subsection{A complex lens system}
\label{subsec:2D}

\begin{figure*}
\begin{tabular}{cc}
\includegraphics[scale=0.55]{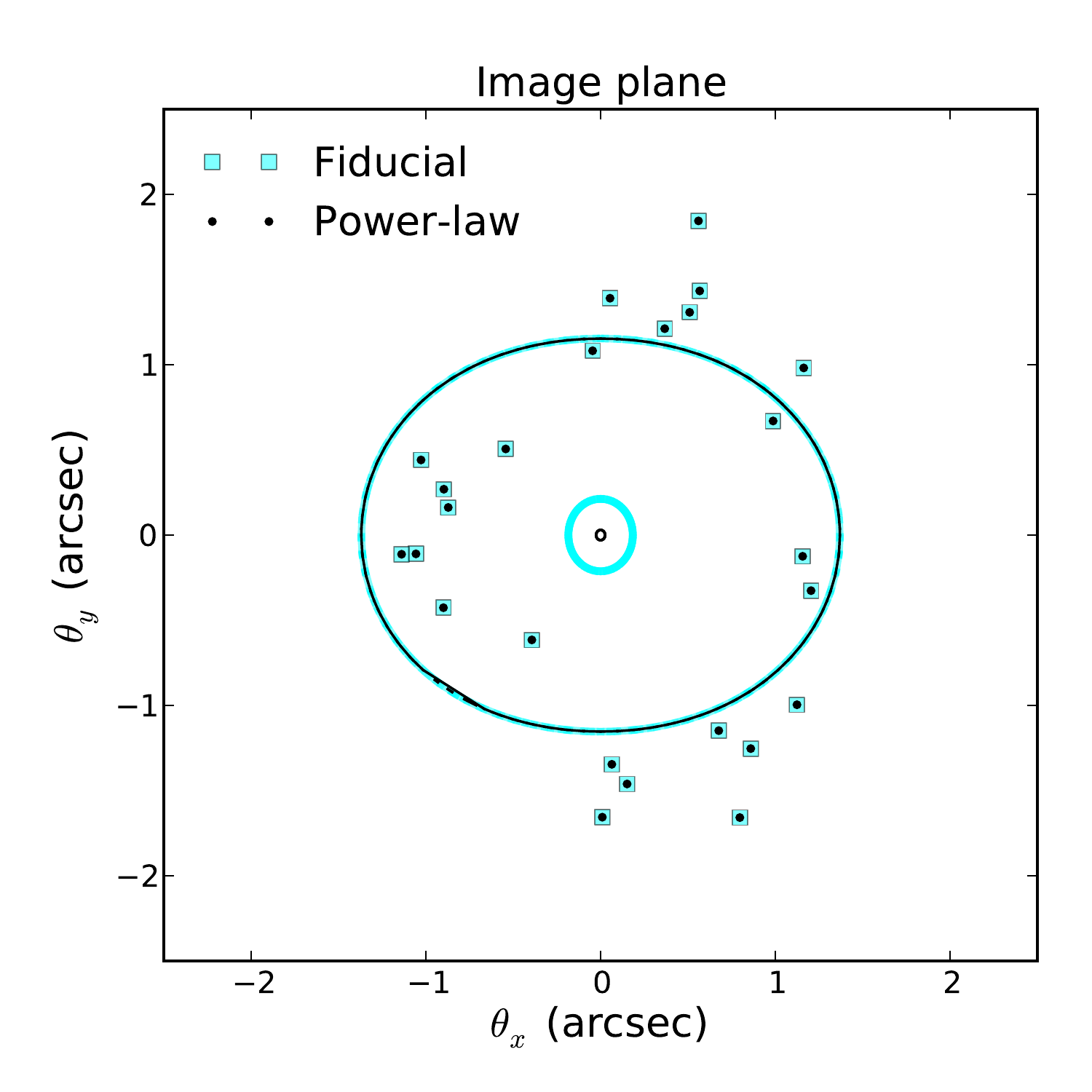} & \includegraphics[scale=0.55]{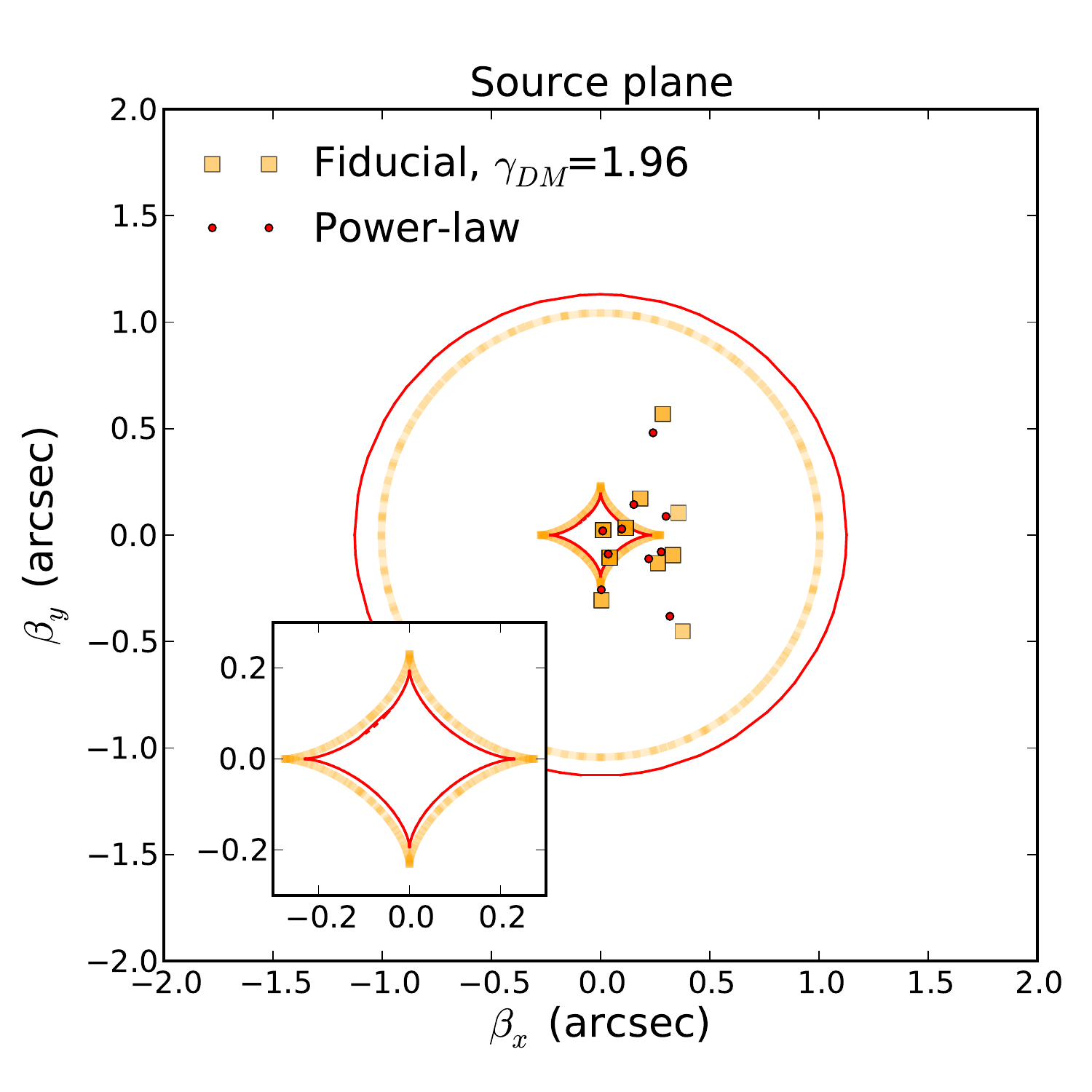} \\
\end{tabular}
\caption{Example of a complex lensed image system obtained from
  lensing by a composite mass distribution (Hernquist+gNFW) and well
  reproduced with a power-law mass distribution. {\it {Left}}: Lensed
  image positions and critical curves for the fiducial composite model
  (light cyan) and for the power-law model (black). {\it {Right}}:
  Corresponding caustics and source positions for the fiducial
  composite model (light orange) and for the power-law model
  (red). The small inset panel provides a zoom of the inner
  caustic }
\label{fig:caustics}
\end{figure*}

Our toy data set indicates that degeneracies between lens models could
be a serious limitation to accurate lensing analysis. This model,
however, provides a highly simplified view of a lens system. Real lens
systems show more complex image configurations that need to be fitted
with mass models, within the observational accuracy of positions.  We
now routinely measure lensed image positions to a few milli-arcsec
accuracy and observe Einstein ring-like features.{\footnote{The
    Einstein-Chowlson ring originally refers to the image of the
    source in case of perfect alignment between the source, lens and
    observer. We follow the common lensing jargon and use this
    terminology to describe ring-like highly deformed lensed images of
    an extended source. }} It has been shown that those observational
constraints were crucial to disentangle lens models
\citep[e.g.][]{Kochanek2001, Cohn2001, Wucknitz2004, Suyu2006,
  Sluse2008, Suyu2010b, Sluse2012a, Oguri2013} and we may think that
this is sufficient to break the degeneracy outlined in the previous
section (except, of course, the MSD). In order to investigate this
question, we have simulated various ensembles of arbitrarily complex
sources composed of 10 point-like features paving the source plane,
and lensed them with a plausible lensing galaxy. We have then used
power-law density profiles to fit the positions of the mock lensed
images, and compared the retrieved flux ratios and time delays with
the simulated ones.

A plausible model for a lensing galaxy has to include at least one
component for the baryons and one for the dark matter halo. Following
\cite{Courbin2011} in their study of the lensed system HE0435--1223, we
have decided to model the stellar component with a Hernquist mass
distribution and the dark matter halo with a generalized NFW
profile. The density profile of the Hernquist
(\citeyear{Hernquist1990}) profile is expressed as
\be
\rho_{\star}(r) = \frac{\rho_{\star, c}}
{(r/r_{\star})(1+r/r_{\star})^3}\;, 
\ee
where $\rho_{\star, c}$ is a characteristic density and
$r_{\star}$ the scale radius of the profile. The generalized NFW profile \citep{NFW96} is
\be
\rho_{h}(r) = \frac{\rho_{\rm h, c}}
{(r/r_{\rm s})^{\gamma_{\rm DM}}
[1+(r/r_{\rm s})^2]^{(3-\gamma_{\rm DM})/2}}\;, 
\ee
where $r_{\rm s}$ is the scale radius, $\gamma_{\rm DM}$ the inner
slope of the density profile and $\rho_{\rm h, c}$ is a characteristic
density. 

This particular choice of the mass distribution for the composite lens
is to some extend arbitrary, but is sufficient for our
aims. \cite{Courbin2011} generated an ensemble of galaxies which were
dynamically stable, and had a stellar velocity dispersion compatible
with the one observed in HE0435--1223. We used a subsample of those
galaxies to create a set of about 1000 mock lensing galaxies. For
simplicity, we kept the mass distribution spherically symmetric, and
added an external shear $(\gamma, \theta_\gamma)=(0.1,
90^{\circ})$. The Einstein radius was kept fixed at $\theta_{\rm
  E}=1\arcsecf25 \pm 0\arcsecf05$. We also fixed the scale radii of the two mass
components at $\theta_{\rm s}=r_{\rm s}/D_{\rm d}\sim 4\, \theta_{\rm
  E}$, $\theta_{\star}=r_\star/D_{\rm d} \sim 0.8 \, \theta_{\rm E}$);
the two remaining free parameters of the model are then the ratio of
the normalization of the two mass components and the inner dark matter
slope $\gamma_{\rm DM}$.  

For each lens model, we have randomly
generated 10 point-like components distributed in the source plane
with a uniform probability distribution between the lens center
$\beta=0$ and $\beta_{\rm max}$. We have calculated the positions of the
lensed images using the public lens modeling code \texttt{lensmodel}
(v1.99) developed by Keeton (\citeyear{Keeton2001}), and assigned them
an uncertainty of 0.004 arcsec on the position of each lensed image,
representative of the best existing observational constraints. A
better astrometric accuracy is sometimes achieved for radio or optical
data of a few compact lensed images, but those positions often reveal
astrometric perturbations produced by substructures close in
projection to the lensed images and not accounted for by the smooth lens
models. Those synthetic lensed images should also be equivalent to the
constraints provided by a smooth extended image, such as an
Einstein ring observed with the
Hubble Space Telescope. Those synthetic data are then fitted with a
power-law mass distribution
\be
\kappa(\theta) = \frac{1}{2}\,
\frac{b^{\gamma'-1}}{(\theta^2+\theta^2_{\rm c})^{(\gamma'-1)/2}} \;
\label{equ:PL}
\ee
plus external shear, where $b$ is a normalization factor, $\theta_{\rm
  c}$ is the core radius of the profile (assumed to be arbitrarily
small, unless otherwise stated) and $\gamma'$ is the logarithmic slope
of the 3D density profile [i.e. $\kappa(\theta) \to
\theta^{1-\gamma'}$ and $\rho(\theta) \to \theta^{-\gamma'}$], with
$\gamma'=2$ for an isothermal profile.

We found that the power-law mass distribution could in general
reproduce the mock lensed images extremely well (i.e. with a reduced
$\chi^2 \leq 1$). We show in Fig.~\ref{fig:caustics} one particular
example for which the lensed images generated with the composite model
are perfectly reproduced with an almost isothermal lens. We clearly
see that the two models have a nearly identical outer (tangential)
critical curve. The inner critical curves are more different, but this
does not come as a surprise since our power-law model was chosen to be
nearly singular while our composite model has a relatively flat
core. We discuss this effect hereafter.

Because the ability to fit a power-law model to the simulated lensed
images did not depend on the parameters explored by our composite
model (i.e. variable baryonic fraction $f_{\rm b}$ and $\gamma_{\rm
  DM}$), we focus on the fiducial example presented in
Fig.~\ref{fig:caustics} in the following. The fiducial model has an
average projected baryon fraction $\ave {f_{\rm b} (<\theta_{\rm E})} \sim 0.4$,
and $\gamma_{\rm DM}=1.96$. The lensed images generated with this model
are fitted by a power law model with $\gamma'\sim $1.98. The
approximate isothermality of the fitted power law model is not a
generic result. For our ensemble of mock lenses, the fitted
logarithmic slope of the power law was found in the range
$\gamma' \in$ [1.55, 2.01]. The observed trend to fit model
shallower than isothermal is likely incidental and is related to our
particular choice of $r_{\rm s}$ and $r_\star$. As we show hereafter, the
slope of the fitted power law depends on the surface density close to
the galaxy center.

\begin{figure}
\includegraphics[width=\hsize]{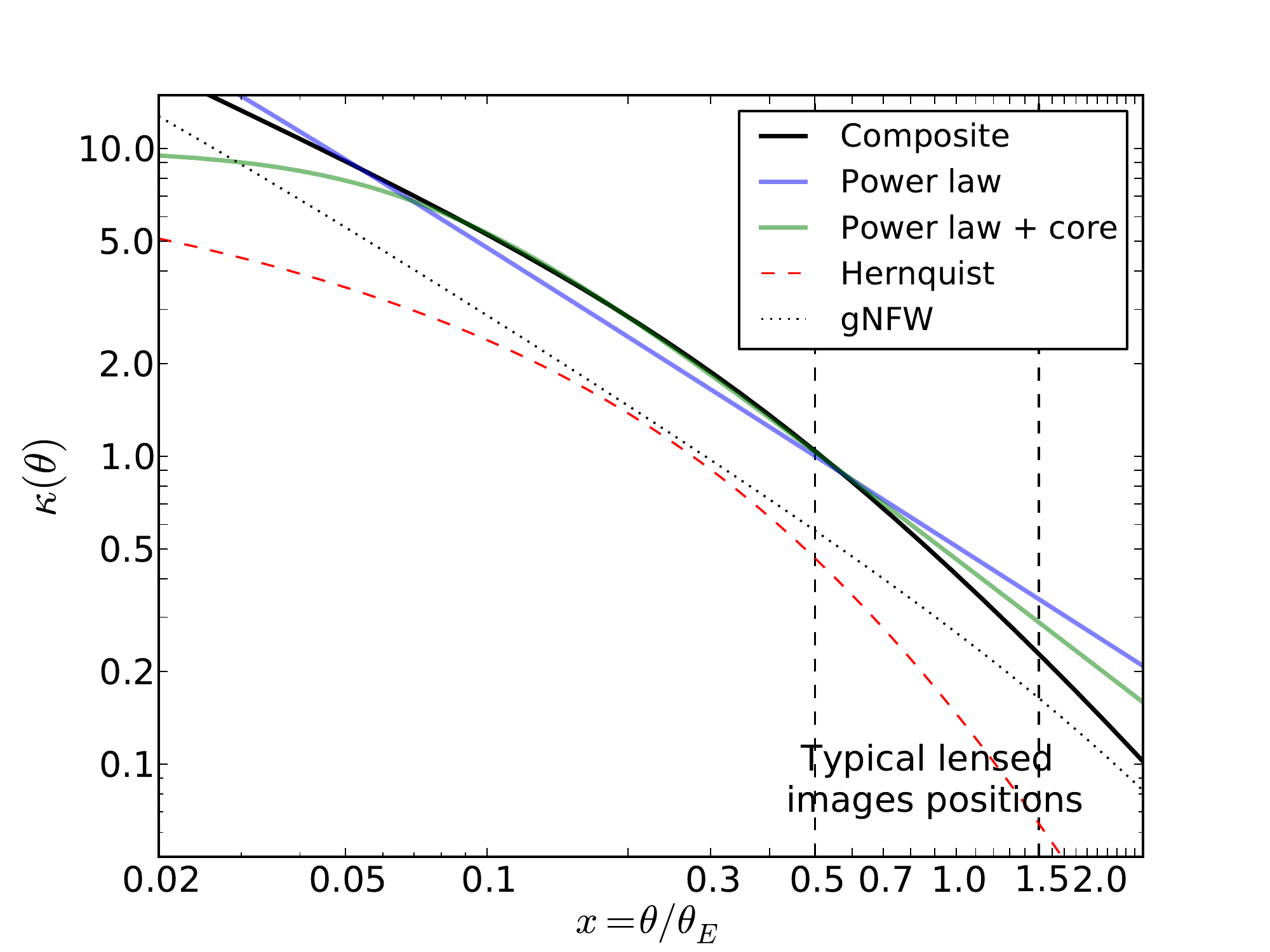}
\caption{Surface mass density of our fiducial model composed of
  a Hernquist (dashed red) + generalized NFW (dotted black) model. Two
  power-law models which reproduce equally well an ensemble of lensed
  images whose positions range typically in $x \in [0.5, 1.5]$
  (vertical dashed lines) are shown. The solid blue curve is for a
  singular power-law model, whereas the green curve is for a power law with
  a core (see Sect.~\ref{subsec:2D})} 
\label{fig:density}
\end{figure}

The projected density profiles of the fiducial lens and of the fitted
power-law profile are shown in Fig.~\ref{fig:density}. 
The slope and the surface density of the two models are very
different. However, both models reproduce the same set of 26 lensed
images (Fig.~\ref{fig:caustics}). Comparing the source morphology
found for the fiducial and the power-law model, we see that the latter
is simply a scaled-down version of the one found for the fiducial
model, a behavior similar to what would be obtained with a mass-sheet
transformation.  In fact, the fitted power-law model (PL) is, to a
good approximation, a mass-sheet transformed version of the fiducial
model (fid), with source positions $\vc\beta_{\rm PL} =
\vc\beta_{\rm fid}/\lambda$, and magnifications $\mu_{\rm PL} \sim
\lambda^2\,\mu_{\rm fid}$, with $\lambda \sim 1.19$. For a MST, the
time delays are also scaled as $\tau_{\rm PL} \sim \tau_{\rm
  fid}/\lambda$. This scaling of the time delays is verified in our
fiducial case but not for the ensemble of mock lenses. The simple
scaling of the source positions and magnifications implies that having
a source which is extended instead of composed of many point-like
structures would not help to break the degeneracy: the power-law and
the fiducial model will lens two sources linearly scaled by a factor
$\lambda$ into an almost identical Einstein ring.

\begin{figure}
\includegraphics[width=\hsize]{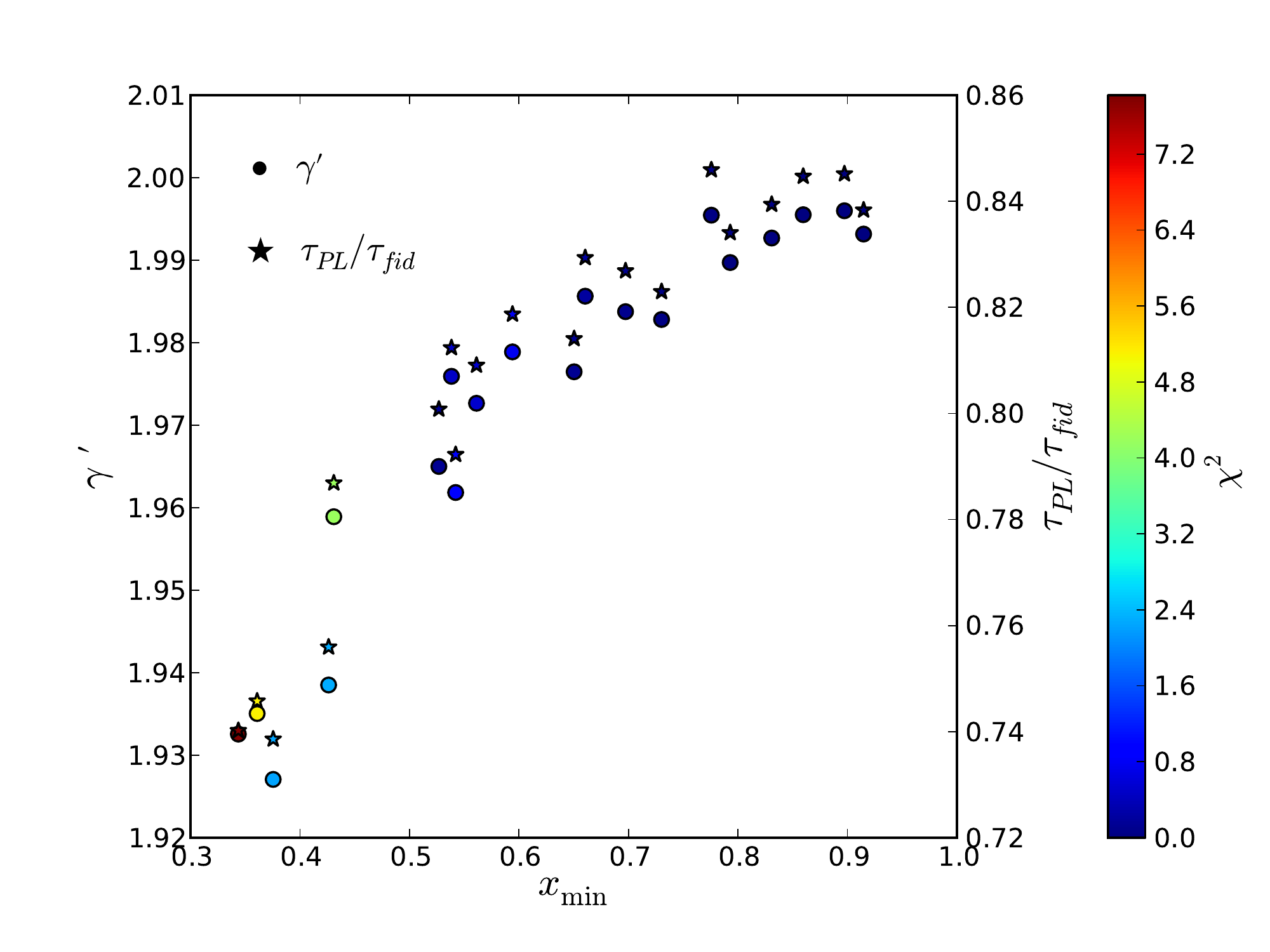}
\caption{Result of the fit of the (singular) power-law model to the
  composite model for 20 different sets of complex sources. For each
  realization of the complex source, we show the value of the
  logarithmic slope $\gamma'$ (circles) of the fitted power law as a
  function of the position of the inner-most lensed image in the
  system ($x_{\rm {\rm min}}$). The time-delay ratio is also shown
  (star-like symbols). The color code shows the (non-reduced) $\chi^2$
  obtained for the fit of the power-law model; the typical number of
  degrees of freedom is $\sim 30$}
\label{fig:chi2}
\end{figure}

In order to investigate whether some particular source configurations
could break the degeneracy, we have generated 20 new sets of complex
sources and lensed them with our fiducial model. Figure~\ref{fig:chi2}
displays the logarithmic slope of the fitted power law as a function
of the position of the innermost lensed image $x_{\rm min}$ and
colored based on the value of the $\chi^2$. We see that the power-law
model provides an excellent representation of the mock data over a
large range of image positions. However, the slope of the fitted power
law decreases by a few percents with the increase of the width of the
annulus containing the lensed images. Interestingly, the power-law
model fits less accurately the mock data when images at a radius $x <
x^{\rm cut}_{\rm {min}}$ are observed. For our fiducial model, this
happened when $x^{\rm cut}_{\rm {min}} \sim
0.5$. Figure~\ref{fig:caustics} indicates that this situation occurs
because the singular power-law model does not reproduce the inner
critical curve.

In order to better match the inner critical curve, we have fitted 
the lensed images using a power-law model with a freely varying
core $\theta_c$ (see Eq.\ts\ref{equ:PL}). We generated 80
different sets of sources with 10 components each to produce images
very close/far from the center of the galaxy. This model allows us to
obtain a nearly perfect fit ($\chi^2$ < 0.5) of all the 80 images
configurations (down to $x_{\rm min} \sim 0.2$). The power law
retrieved with that model has a relatively large core radius
($\theta_{\rm c} \sim 0\arcsecf13$) and a steeper density profile
($\gamma' \sim 2.3$) which reproduces almost perfectly the one of the
fiducial model between $0.15 < x < 0.7$ (Fig.~\ref{fig:density}). Like
in the case of the singular power law, the fitted value of $\gamma'$
depends on the range over which lensed images are observed. Again, the
sources obtained with the two models (PL and fiducial) are almost identical
up to an isotropic scaling factor $\lambda$. The time delay ratios are,
however, not conserved perfectly between the two models, with
deviations reaching several tens of percents for some image
pairs. This shows that the transformation between the two lens models
is not simply a MST.

{{We can now ask the question whether the power law models are also compatible with the velocity
    dispersion of the lens. The composite model we used has a
    velocity dispersion $\sigma=261\,$km/s, as measured within a
    $1^{\prime\prime}$ slit. Following the combined lensing+virial
    formalism introduced by \cite{Agnello2013}, we have estimated that
    the luminosity-weighted line of sight velocity dispersion of the
    power-law model, measured within a $1^{\prime\prime}$ aperture, is
    $\sigma_{\rm {los}, L(R<1 ^{\prime\prime})} = 262\,$km/s, in
    perfect agreement with the value found for the fiducial model. The
    velocity dispersion of the power law with a finite core,
    characterized by $\gamma^{\prime}=2.3$, is more difficult to
    assess. Following the same prescription as above, we find
    $\sigma_{\rm{los}, L(R<1 ^{\prime\prime})} = 375\,$km/s. However,
    this estimate does not account for the presence of a finite core in
    the center of the profile. To estimate the impact of the core, we
    use the results of Keeton \& Kochanek (\citeyear{Keeton1998b}; see
    also Dutton et al. \citeyear{Dutton2011}), to derive the circular
    velocity of an isothermal profile in the presence of a
    core. Following that formalism, we find for a core radius of
    $\theta_c=0 \arcsecf 13$ that the circular velocity is reduced by
    $\sim 30 \%$ at a radius of half the slit width
    (i.e. $0\arcsecf5$).  Assuming that this scaling of the circular
    velocity is also valid for non-isothermal profiles, and knowing
    that the velocity dispersion is proportional to the circular
    velocity \citep{Padmanabhan2004, Capellari2013}, we calculate the
    velocity dispersion of the power law model with a core,
    $\sigma_{\rm{los}, L(R<1 ^{\prime\prime})} = 288\,$km/s. Hence,
    the power-law models which are degenerate with our composite
    fiducial model also have a velocity dispersion compatible with that
    one if velocity dispersion uncertainties in agreement with typical
    observations (i.e. about 10\%) are assumed. We may be careful in
    generalizing this result too much, since our fiducial composite
    model and power-law models are not universal representations of
    the density profile of lensing galaxies. The velocity dispersion
    will always reduce the family of degenerate models reproducing the
    data, but our illustrative example shows that models predicting
    very different time delays can be found. }}

The above examples show that even for a non-circular lens potential
there is significant freedom in choosing the lens model{{, even in
    presence of a velocity dispersion measurement}}.  This freedom is
due to the fact that our composite mass model transforms into an
approximate local power law by means of a mass sheet-like
transformation. It is important to realize that this transformation is
formally unrelated to the external convergence from the environment
(which we discuss in the next section). It exists because of the
freedom one has in modifying the gravitational potential in regions
not probed by the lensed images (i.e. for $x < x_{\rm min}$ and $x>
x_{\rm max}$). The freedom we have is however limited because the mass
interior to the Einstein radius cannot change between the two
models. A measurement of the time delays between lensed images,
assuming a value of $H_0$ determined by other methods, can break this
degeneracy since $\tau$ is mostly sensitive to the slope of the mass
profile near the Einstein radius \citep{Gorenstein1988, KOC2002}. The
observation of a second source at a different redshift should also
break that degeneracy. {{Indeed, in case of multiple sources, a
    set of equations similar to (\ref{eq:MSD}) has to be valid for
    each different source redshift, hence different distance ratios
    $D_{\rm ds}/D_{\rm s}$. As explained in details in Saha
    (\citeyear{Saha2000}, Sect.\ 2.2), \cite{Bradac2004b}, or
    Liesenborgs \& De Rijcke (\citeyear{Liesenborgs2012}, Sect.\ 4),
    this effectively breaks the MSD.}} This implies that strong
lensing studies by galaxy clusters, where sources at multiple
redshifts are commonly observed, should be less impacted by that
degeneracy. However, compact substructures in the potential, producing
monopole-like degeneracies, may be a concern \citep{Liesenborgs2009,
  Liesenborgs2012}. For galaxy-scale lenses, sources at multiple
redshifts are rarely observed, but a few systems with double Einstein
rings are known{\footnote{{{The double-source nature of the lens
        SDSSJ0924+0219 reported in \cite{Eigenbrod2006a} is more
        speculative than for the system SDSSJ0946+1006 -- also known
        as the ``Jackpot'' -- discovered by \cite{Gavazzi2008} and
        further studied by \cite{Sonnenfeld2012}. }}}}
\citep{Eigenbrod2006a, Gavazzi2008}. We can also naively think that
the observation of a central image would be of great relevance to
break the degeneracy, but the magnification of the latter image
constrains only the very inner core of the lensing galaxy and may be
unrelated to the global shape of the mass distribution.

{After the first version of this paper was published on the arXiv,
  \cite{Suyu2013new} modeled the time-delay lens system J1131$-$1231
  with a composite model, using the observed light profile of the lens
  to fix the baryonic component, and a standard NFW-profile for the dark
  matter. This model yielded a value for $\tau$ which is good agreement with
  their earlier analysis based on a power-law mass model.  This result
  shows that the same data can be reproduced with more than one class
  of models. However, over the radial range where lensing data are
  available, the mass profile of the composite lens and the
  best-fitting power-law model essentially coincide.  Therefore, the
  new model does not probe the freedom of the lens model offered by
  the MST. It is possible that this agreement is a coincidence,
  originating from the specific shape of the luminosity profile and
  the choice of an inner slope of the dark matter profile of $-1$. As
  mentioned before, this slope is obtained from dark matter-only
  simulations, but is most likely altered once baryon cooling comes
  into play \citep[see, e.g.,][and references therein]{Scanna12}.}

\section{The external convergence}
\label{sec:kappaext}

The original motivation for Falco et al.\ (\citeyear{FGS85}) to
consider the MSD was related to the first gravitational lens system
0957+561, where the main lens is embedded in a group or low-mass
cluster of galaxies. The unexpectedly large angular separation of
$\sim 6''$ in this system hinted at a substantial influence of the
mass distribution of the cluster on the lensing properties of the
lens. Neglecting shear, the prime effect of the cluster is to yield an
additional surface mass density at the location of the main lens
galaxy, which boosts the angular splitting. Falco et al.\ pointed out
that there is a trade-off between the mass parameters of the main
lensing galaxy and the amplitude of this external convergence, which
is described by the MST (\ref{eq:MSD}). This physical interpretation
yielded the name of the transformation.

Based on that, one frequently finds the MST written in the form
\be
\kappa(\vc\theta) \to \rund{1-\kappa_{\rm ext}}\kappa(\vc\theta)
+\kappa_{\rm ext}\;,
\label{eq:MSD2}
\ee
which obviously is fully equivalent to (\ref{eq:MSD}). However,
the way (\ref{eq:MSD2}) is written suggests a physical interpretation
of the MST: if the original mass profile $\kappa(\vc\theta)$ is chosen
such that $\kappa\to 0$ as $|\vc\theta|\to\infty$, then 
the parameter $\kappa_{\rm ext}$ is interpreted as the convergence far
from the lens (e.g., the convergence contributed by a large-scale
matter inhomogeneity in the direction of the main lens, such as a
cluster in which the lens is embedded). In this case, one may
investigate the line-of-sight to the lens to find indications for an
over- (or under)density of galaxies, which can then be used as a proxy
for the corresponding $\kappa_{\rm ext}$. 
In contrast, in its form
(\ref{eq:MSD}) the transformation is of purely mathematical nature,
without any interpretation of $\lambda$. 

We believe that the interpretation of (\ref{eq:MSD2}) is
misleading. Suppose some mass distribution $\kappa(\vc\theta)$
provides a good model fit to the lensing data, and suppose that this
analytical model (such as an isothermal ellipse) is chosen as to
vanish for large radii. Then for any $\lambda\ne 1$, the asymptotic
value of $\kappa_\lambda(\vc\theta)$ for large separations is
$1-\lambda$, i.e., in general different from zero. However, this does
{\it not} imply that the lens is embedded in an external convergence
with $\kappa_{\rm ext}=1-\lambda$: the strongly lensed images are
located close to the center of the lens, typically within two Einstein
radii, i.e., in regions where $\kappa\gtrsim 0.3$. Assuming elliptical
symmetry, the mass profile outside this region does not affect the
lensing properties, and is therefore not constrained. We note however
that in case of non-elliptical symmetry, due to a gradient of
convergence produced by nearby companion galaxies, or a group or
cluster, lensing observables will be affected. There are many examples
where this happens, including many time-delay lenses.

Assuming then that the analytic profile
$\kappa(\vc\theta)$ is a good description of the {\it global} mass
profile, just because it provides a good fit to the lensing
constraints, is a strong extrapolation, given that we have no good
physical model for the density profile of galaxies -- as mentioned
before, we do not understand why the sum of the baryonic density and
the dark matter distribution superpose to an almost isothermal profile
at radii close to the Einstein radius. 

\begin{figure}
\includegraphics[width=\hsize]{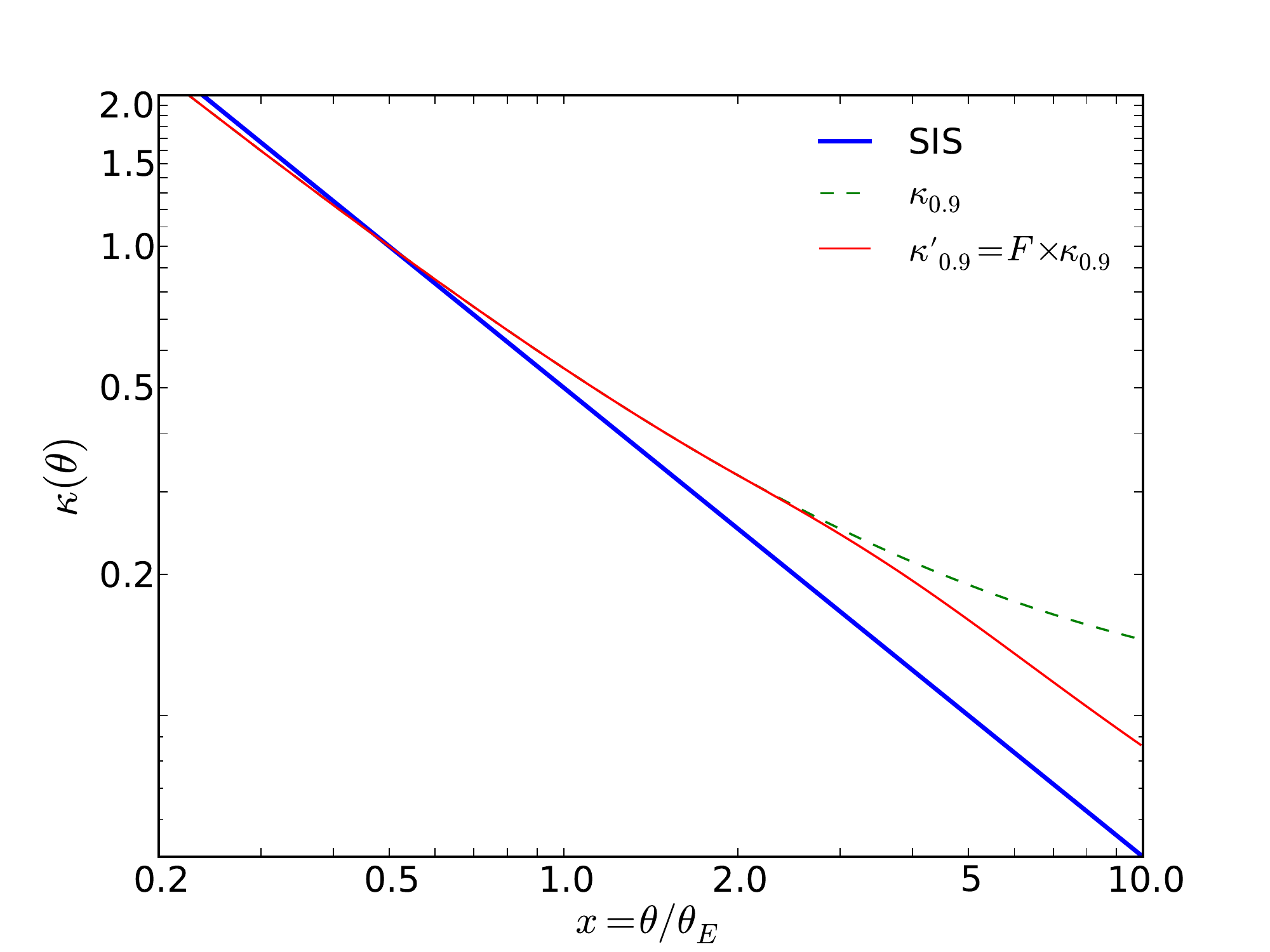}
\caption{The solid blue curve shows the isothermal density profile,
  the dashed curve shows the mass-sheet transformed profile
  $\kappa_\lambda$ with
  $\lambda=0.9$, and the thin solid red curve shows $\kappa_{0.}$,
  modified by multiplication with a smooth function $F$ as described
  in the text}
\label{fig:4}
\end{figure}

Mathematically, we can define a function $F(\vc\theta)$ which is unity
for $|\vc\theta|\lesssim 2\theta_{\rm E}$, and then decreases smoothly
to zero for larger radii. Then, $\kappa'_\lambda=\kappa_\lambda F$ is
unchanged in the inner regions of the lens (provided $F$ shares the
same symmetries as $\kappa$, e.g., being constant on confocal
ellipses) and decreases to 0 at large separation $|\vc\theta|$, with
no relation to an external convergence. Conversely, we can slightly
redefine $F$ such that $\kappa'_\lambda$ decreases to any desired
value of $\kappa_{\rm ext}$ for large radii. Fig.\ts\ref{fig:4}
illustrates this with a simple example for such a transformation.
Hence, studying the environment of a lens may be of limited use for breaking the MSD.

\section{Conclusions}

The mass-sheet degeneracy is the largest obstacle in gravitational
lensing to obtain model-independent accurate quantitative
results. Apart from the mass inside the Einstein radius, the masses
inside all other radii are affected by the MSD, as is the (slope of
the) radial mass profile.  One instructive example of that statement
can be found in Fig.\,5 of \cite{Coe2012}, where different
methods for obtaining the density profile in the inner region of the
lensing cluster Abell\ts 2261 were employed; the resulting mass
profiles are quite different, despite the fact that several strong
lensing systems are identified in this cluster. For most galaxy-scale
lenses, the observational situation is often worse, with only one source
being multiply imaged in general.

The MST affects the product $H_0\,\Delta t$ of the Hubble constant and
the time delay, and hence the determination of $H_0$ from lens
systems. In this paper, we have pointed out a number of issues related
to mass models and the MSD; in particular, we question the commonly
made assumption that the density profile is a global power law, an
assumption which implicitly breaks the MSD. Given our lack of
understanding of the conspiracy that yields a mean slope close to an
isothermal profile from a superposition of a S\'ersic profile, which
describes the luminous matter in galaxies, and that of the dark matter
profile, we believe that it is timely to investigate the impact of the
single power-law assumption on lensing results.  This is particularly
true close to the region of strong lensing features, where the
observed fraction of dark matter is about one half, i.e., most likely
the transition region from baryon-dominance to dark matter-dominance
in lens galaxies. Using plausible lens models, we have shown that
composite models made of a S\'ersic (or Hernquist)+dark matter profile
could be transformed through a MST-like transformation into a mass
distribution close to power law over the range of radii constrained by
lensed images. This probably explains why galaxy-lensing data may
generally be well fitted with power law mass distributions despite the
more complex intrinsic mass distribution of the lens.  On the other
hand, the MST (\ref{eq:MSD}) transforms a power law into a profile
which resembles to good approximation a power law, but not an exact
one. Preference of an exact power law therefore seems to some degree
arbitrary.

We also question the physical interpretation of the MST to be directly 
related to an external convergence, since this implies a 
extrapolation of the density profile well beyond the region where the
mass profile is constrained by lensing information. Again, due to the
lack of accurate physical models for the density distribution in
galaxies, such an extrapolation may not be appropriate. In particular,
the frequently employed isothermal profile must break down at large
radii, since the {{dark matter profile}} is steeper than
isothermal for large $r$, and also because the total mass diverges
linearly. 

Together, we therefore question whether the determination of the
Hubble constant from gravitational lensing indeed is as accurate as
sometimes claimed in the literature. Whereas the freedom offered by
the MST (\ref{eq:MSD}) cannot be stretched arbitrarily -- values of
$\lambda$ too different from unity may indeed lead to unphysical mass
profiles, and in particular violate constraints from the measured
stellar velocity dispersion of the lens galaxy, some $\sim 10\%$
deviations (or even larger) of $\lambda$ from unity seem quite
plausible. {{The Hubble constant as determined with free-form lens
    modeling \citep[e.g.][]{Saha2006, Coles2008, Paraficz2010} may be
    less prone to the MSD because free-form lens models naturally
    explore a large variety of mass distributions. Enlarging the
    parameter space may, however, not guarantee an unbiased
    estimate of $H_0$. Indeed, despite some priors on the variation of
    the potential with radius \citep{Saha2004}, there is no security
    that unphysical mass distributions (e.g. dynamically unstable) are
    not part of the ensemble of valid lens models. The consistency
    between $H_0$ as derived with pixelated mass models and the value
    published by Planck indicates that the current method works
    reasonably well, but we think that breaking the MSD with
    free-form lens models is not yet enough controlled to derive an
    {\it {accurate}} value of $H_0$.}}

{{Whatever the lens modeling approach, the impact of the MST on
    the value of $H_0$ from gravitational lensing has to be quantified
    carefully. This is particularly true because the resulting effect
    on $H_0$ may not be statistical, and thus not average out when
    considering samples of lens systems. If the true mass profile in
    the strong lensing region of galaxies is systematically curved
    (say convex, in the transition region between baryon dominance and
    dark matter dominance), the power-law assumption will yield a
    systematic bias on the estimates of $H_0$. This is because the
    sign of $(1-\lambda)$ determines the sign of curvature a
    mass-sheet transformed power-law model will attain.}} 
  \cite{Fadely2010}, in their in-depth analysis of the lens system
    Q0957+561 using two-component mass models and constraints from
    weak lensing, have investigated how their estimate of $H_0$ was
    limited by the MSD.

On the other hand, a better understanding of galaxy formation
    and the physics of galaxies may allow one to reduce the freedom in
    lens models. The apparent conspiracy for lensing galaxies to
show profiles close to isothermal around one Einstein radius, or even
at larger radii \citep[e.g.][]{Gavazzi2007, Koopmans2009,
  vandeven2009, Humpfrey2010, Dutton2013}, could be an indication that
the systematic errors introduced by the choice of an isothermal model
are small or could be quantified based on some observational
properties of the galaxies. The study of mock lens galaxies based on
parametrized density distributions \citep{vandeven2009} or
hydrodynamical simulations may enable one to estimate quantitatively
how large this effect is on real galaxies, and to see whether
systematic errors introduced by the use of a particular type of lens
models cannot be calibrated.

From the above discussion, it is also clear that if accurate
measurements of the Hubble constant are obtained with an independent
method, lenses with measured time delays are the best suited ones to
probe the density slope in the inner part of galaxies \citep{Read2007,
  Fadely2010}. The modeling of such systems may enable one to
investigate deviations from simple power-law profiles, and, together
with a determination of the baryonic mass from the observed light
profile of the lens galaxy and its velocity dispersion, to learn about
the shape of the dark matter profile in the inner region of galaxy
halos. This indeed makes time delay lenses a unique tool for studying
the inner parts of dark matter halos in galaxies.

\begin{acknowledgements}
  We thank Sherry Suyu for numerous and sometimes controversial
  discussions which have inspired this work. Olaf Wucknitz, Yves
  Revaz, Chris Kochanek, Andrea Macci{\`o} are also acknowledged for
  useful discussions. We thank the anonymous referee for his
  thoughtful comments on the manuscript.  Part of this work was
  supported by the German \emph{Deut\-sche
    For\-schungs\-ge\-mein\-schaft, DFG\/} project number SL172/1-1.
\end{acknowledgements}

\bibliographystyle{aa}
\bibliography{MSDbib}



\end{document}